# Angular dependent magnetothermopower of α-(ET)$_2$KHg(SCN)$_4$

D. Krstovska[*], E. Steven, E. S. Choi, and J. S. Brooks[1]

*National High Magnetic Field Laboratory, Florida State University, 1800 East Paul Dirac Drive, Tallahassee, Florida 32310, USA*

**Abstract.** The magnetic field and angular dependences of the thermopower and Nernst effect of the quasi-two-dimensional organic conductor α-(ET)$_2$KHg(SCN)$_4$ are experimentally measured at temperatures below (4 K) and above (9 K) the transition temperature to fields of 31 T. In addition, a theoretical model which involves a magnetic breakdown effect between the q1D and q2D bands is proposed in order to simulate the data. Analysis of the background components of the thermopower and Nernst effect imply that at low temperatures, in the CDW state, the properties of α-(ET)$_2$KHg(SCN)$_4$ are determined mostly by the orbits on the new open Fermi sheets. Quantum oscillations observed in the both thermoelectric effects, at fields above 8 T, originate only from the $\alpha$ orbit.

[*] Permanent address: *Ss. Cyril and Methodius University, 1000 Skopje, Macedonia*
[1] Corresponding Author

## 1. INTRODUCTION

The quasi-two-dimensional organic conductor $\alpha$-$(ET)_2KHg(SCN)_4$ has been extensively studied over the past decade because of the highly unusual behavior under magnetic fields. This compound is a member of the isostructural family $\alpha$-$(ET)_2MHg(XCN)_4$, where M = K, Tl, Rb, or $NH_4$ and X = S, Se. The Fermi surface (FS) of this salt contains both a q2D cylinder and a pair of weakly warped open q1D sheets. The q1D and q2D bands are separated by a substantial gap near the Fermi level. At $T_p = 8$ K, the system undergoes a phase transition from a metallic state into a specific low-temperature phase (LTP), displaying numerous anomalies in a magnetic field. The formation of the LTP is associated with the nesting instability of the q1D band resulting in a charge-density wave (CDW). The high-field regime is defined by the so-called kink field $B_k$ at which the zero-field state $CDW_0$ is transformed into the $CDW_x$ state with a field-dependent wave vector. The energy gap is strongly reduced in the $CDW_x$ state in comparison to that in the $CDW_0$, so that the cyclotron orbits on the cylindrical part of the FS are considered to be the same as in the normal metallic state.

Over the years, several models have been proposed to reveal the origin of the angle dependent magnetoresistance oscillations (AMROs) of LTP in $\alpha$-$(ET)_2KHg(SCN)_4$. AMROs have been also studied by assuming that the LTP is unconventional charge density wave (UCDW) [1]. This model involves the quantization of the energy spectrum of the q1D charge carriers group in a presence of a magnetic field [2]. However, the origin of the novel AMROs as well as suggested UCDW in $\alpha$-$(ET)_2KHg(SCN)_4$ is still to be confirmed. Recently, the origin of the anomalous AMROs in the LTP was additionally investigated and discussed through the temperature dependence of AMROs [3]. At 2 K in the CDW phase, there exists the anomalous Lebed resonance–like pattern (local maxima in the interlayer magneto-conductivity), the amplitude of which is modulated by Danner-Kang-Chaikin (DKC) oscillations when the applied magnetic field is close to the q2D conducting plane. In contrast, at 7 K just below the transition temperature $T_p$, local minima, i.e. Kajita oscillations originating from the cylindrical FS, superposed on the anomalous Lebed resonance were observed. This indicates that the magnetic breakdown (MB) effects are crucial for the anomalous AMROs in the CDW phase of $\alpha$-(ET)$_2$KHg(SCN)$_4$.

In the recent years, the magnetothermopower (MTEP) of organic conductors with q1D and q2D electron energy spectrum has been both experimentally [4, 5] and theoretically [6-11] studied. The investigation of MTEP particularly in quantized magnetic field opens new possibilities and can give many interesting data not only for the structure of the energy spectrum and the scattering processes. It may also be used to explain phonon interactions in solids as well as the acoustic energy absorption at high frequencies, where direct experimental measuring of the acoustic absorption coefficient to date is practically impossible. MTEP measurements yield information about both the thermodynamic and transport properties of carriers. Advantages of thermopower include the zero-current nature of the measurement, and its sensitivity to band structure, especially in the case of anisotropic (low dimensional FS) materials. Choi *et al*. [5] have applied a low-frequency MTEP method to investigate the low temperature, magnetic field dependent phase of $\alpha$-(ET)$_2$KHg(SCN)$_4$. The most significant results of their investigation include the study of the onset of MB effects via the MTEP, and the measurement of magnetic quantum oscillations in the TEP and Nernst effect. A significant anisotropy in the MTEP signal as well as large quantum oscillations associated with the Landau quantization has been observed. Later, Dóra *et al*. [12] studied the thermopower and Nernst effect in UCDW in the presence of magnetic field and compared their results with the experimental data obtained by Choi *et al*. They found their results to be consistent with the experimental data. In addition, it has been shown that the obtained Nernst effect is rather large and independent of the imperfect nesting term.

Experiment data on thermopower and Nernst effect angular dependences in q2D organic materials have not been previously reported. Apart of that, a giant resonant Nernst signal was discovered in the metallic state of the quasi one-dimensional organic conductor (TMTSF)$_2$PF$_6$ in tilted magnetic fields near magic angles parallel to crystallographic planes [13, 14]. The giant resonant Nernst voltage has been also observed in the sister compound (TMTSF)$_2$ClO$_4$ [15]. It was shown that this effect appears to be a general feature of these materials, and is also present in the field induced spin density wave phase with even larger amplitude.

Here we report the results of the angular dependence of the MTEP (Seebeck and Nernst effects) in a multiband organic conductor α-(ET)$_2$KHg(SCN)$_4$. Experimental data are obtained for two temperatures, below and above the CDW transition temperature. Due to the small energy gap between the q1D and q2D Fermi sections, MB effects become important and must be taken into account in order to address the MTEP behavior below the CDW transition temperature. A theoretical model of quantum interlayer tunneling demonstrates the behavior qualitatively. Indeed, our results suggest that magnetic breakdown effects are important in explaining the observed features of MTEP. These results corroborate and advance previous results on MTEP in the CDW phase.

## 2. EXPERIMENTAL DETAILS

α-(ET)$_2$KHg(SCN)$_4$ single-crystal sample was synthesized by conventional electrochemical crystallization techniques. A sample from a single batch was used for the experimental condition (heat current parallel to the $b$ axis). The magnetic field $\mathbf{B} = (B\cos\phi\sin\theta, B\sin\phi\sin\theta, B\cos\theta)$, was rotated from the least conducting axis ($b$ axis) closely to the $ba$ plane (fixed $\phi$). Magnetic field and angular measurements of both the thermopower and Nernst signal were carried out by applying sinusoidal heat current along the crystallographic axis $b$ of the single crystal. The resistance was measured by a conventional 4-probe low frequency lock-in technique. The sample was mounted between two quartz blocks, which were heated by sinusoidal heating currents (with an oscillation frequency $f_0$) with a π/2 phase difference. A miniature heater was placed on top of the sample to establish a small temperature gradient along the $b$-axis. The corresponding temperature gradient ($\Delta T$) and the thermal emf with $2f_0$ oscillation frequency were measured. Three pairs of Au wires were attached to the sample along the $b$ axis on the opposite sides of the $ac$-planes of the sample by carbon paste for both resistance measurements and the magnetothermopower measurements. The temperature gradient ($\Delta T$), which was found to be nearly field independent, was set prior to each magnetic field sweep. The procedures of the magnetothermopower measurement method used in this work are detailed elsewhere [4].

## 3. RESULTS AND DISCUSSION

### 3.1. *Theoretical models for the q1D and q2D bands*

In general, the thermopower $S$ can be expressed by the Mott formula [16]

$$S = \frac{\pi^2 k_B^2 T}{3e} \frac{d\ln\sigma(\varepsilon)}{d\varepsilon}\bigg|_{\varepsilon=\mu} \quad (1)$$

where $\sigma(\varepsilon)$ is the energy-dependent electrical conductivity.

When two types of carriers are involved one should sum up the corresponding contributions to obtain the total thermopower as [16]

$$S_{total} = \frac{\sigma_{1D} S_{1D} + \sigma_{2D} S_{2D}}{\sigma_{1D} + \sigma_{2D}} \qquad (2)$$

where $\sigma_{1D}, \sigma_{2D}, S_{1D}$ and $S_{2D}$ are the conductivity and thermopower contributed by carriers from the open-FS sheets and closed hole (electron) pockets, respectively.

The MTEP signal contains two components, the background MTEP which is sensitive to the FS topology, and the oscillatory MTEP which is a manifestation of the Landau quantization of the closed orbit FS. Since the Fermi energies of organic materials are quite low, typically from few tens to one hundred meV, even gaps which are not very small compared to Fermi energy $\varepsilon_F$ can give rise to MB effects at accessible magnetic fields. Honold et al. [17] have reported on a MB in the high field state of α-(ET)$_2$KHg(SCN)$_4$ through the gap of ~23 meV that is of the same order as $\varepsilon_F$ for this compound.

In an attempt to simulate the thermopower and Nernst experimental data for α-(ET)$_2$KHg(SCN)$_4$, we will make use of a quantum model of the interlayer magneto-tunneling for the q1D carriers [3]

$$\sigma_{zz}^{1D}(B,\theta,\phi) = \frac{\sigma_1}{2\pi} \sum_N J_N^2(ck_F \tan\theta) \int_0^{2\pi} dx \frac{1}{1 + (N\omega_c + \frac{\Delta_g}{\hbar}\alpha(\mathbf{B},\mathbf{Q})\sin x)^2 \tau^2} \qquad (3)$$

with

$$\alpha(\mathbf{B},\mathbf{Q}) = 4 e^{-\frac{l^2 Q_x^2}{4}} J_0(k_F Q_x l^2) \sin\{\frac{cQ_x}{2}(\tan\theta\sin\phi + \frac{Q_z}{Q_x})\} \qquad (4)$$

Here, $\sigma_1$ is the electrical conductivity of the q1D carriers in the absence of a magnetic field, $c$ and $l$ are the interlayer and in-plane spacing, $\tau$ is the electron relaxation time, $k_F$ is the Fermi wave number, $\omega_c$ is the cyclotron frequency, $\Delta_g$ is the CDW gap energy, $Q_x$ and $Q_z$ are the in- and out-of-plane components of the wave vector $\mathbf{Q}$ of the CDW, $J_0$ and $J_N$ are the 0-th and N-th order Bessel function.

The contribution to the interlayer conductivity from the q2D carriers can be calculated by using the Boltzmann equation. The following asymptotic expression is obtained within corrections small in the parameters $2t_c/\mu$ and $\gamma = 1/\omega_c\tau$

$$\sigma_{zz}^{2D}(B,\theta,\phi) = \sigma_2 \left\{ J_0^2(\zeta\tan\theta)J_0^2(\xi\tan\theta) + 2h^2(1+\tan^2\theta) \times \right.$$
$$\left. \times \left( J_0^2(\zeta\tan\theta)\sum_{k=1}^{\infty}\frac{J_k^2(\xi\tan\theta)}{2h^2(1+\tan^2\theta)+k^2} + J_0^2(\xi\tan\theta)\sum_{k=1}^{\infty}\frac{J_k^2(\zeta\tan\theta)}{2h^2(1+\tan^2\theta)+k^2} \right) \right\} \quad (5)$$

where $\sigma_2$ is the electrical conductivity of the q2D carriers along the layers in the absence of a magnetic field, $t_c$ is the interlayer transfer integral, $\mu$ is the chemical potential and $\zeta = ck_F\cos\phi$, $\xi = ck_F\sin\phi$, $h = m^*/eB\tau$.

At low enough temperatures where the condition $\omega_c\tau \gg 1$ is satisfied, the quantization of the energy of electrons whose motion in the plane orthogonal to the magnetic field is finite should be taken into account. Their states belong mainly to the q2D sheet of the FS. The q1D part of the FS in the α-phase (BEDT-TTF) salts is an open orbit, and therefore does not undergo Landau quantization in the vicinity of $\mu$. This part of the FS therefore behaves primarily as a carrier reservoir [18] to and from which the carriers can flow in an attempt to minimize the free energy of the system.

The oscillatory part of the thermoelectric coefficients $\alpha_{yz}$ and $\alpha_{zz}$ can be written as a sum of harmonic periodic in the $1/B$ scale [6]

$$\alpha_{yz}^{osc} = \alpha_{zz}^{osc}\sin\phi\tan\theta, \quad \alpha_{zz}^{osc} = \frac{\pi^2 k_B^2 T}{3e}\sigma_0\left(\frac{2}{t_c\hbar\omega_c(\theta)}\right)^{1/2} J_0^2(ck_F\tan\theta) \times$$
$$\times \sum_{k=1}^{\infty} k^{3/2} R_T(k\Lambda) R_S(k) R_D(k) R_W(k) \sin\left(2\pi k\left(\frac{F}{B} - \frac{1}{2}\right)\right). \quad (6)$$

with a fundamental frequency $F = \mu m_c/\hbar e$. $R_T(k\Lambda) = k\Lambda/\sinh(k\Lambda)$, $\Lambda = 2\pi^2 k_B T/\hbar\omega_c \ll 1$, is the temperature damping factor. $R_S(k) = \cos(k\pi g\mu_r/2)$ is the spin-splitting damping factor. $g$ is the conduction electron $g$ factor and $\mu_r = m_c(\theta)/m$ is the angle dependent cyclotron mass $m_c$ normalized to the free electron mass $m$. $R_D = \exp(-\pi k/\omega_c \tau)$ is the Dingle factor and $R_W(k) = J_0(4\pi k t_c/\hbar\omega_c)$ describes the interlayer coupling.

*3.2. Experimental data and comparison with the theory*

In Fig.1, we present the experimental data (solid lines) and theoretically simulated curves (dashed lines) of the magnetic-field dependence of the total TEP (Seebeck effect) signal (containing both the background and oscillatory component) for the temperature gradient along the crystallographic direction $b$ (or coordinate axis $z$), $S_{zz}^b$, at $T = 4$ K, in the CDW state of α-(ET)$_2$KHg(SCN)$_4$. The magnetic field was initially applied along the $b$-axis ($\theta = 0^0$) and then rotated to the $a$-axis (or coordinate axis $y$) for $\theta = 26^0$ and $\theta = 45^0$. The first and the third angle are for the 1D AMRO peaks (Fig. 2a). The second is the angle at which the first dip ($N$=1) in the AMROs is observed, obeying the condition, $\tan\theta_N \sin(\varphi - \varphi_0) = N\Delta_0 + \tan\theta_0$ [19, 20], which is of the same form as generalized Lebed magic angles (LMA) condition. For α-(ET)$_2$KHg(SCN)$_4$ the parameters are estimated to be $\tan\theta_0 = 0.5$, $\Delta_0 = 1.25$, and $\varphi_0 = 27^0$ [1]. When the field is directed at an LMA, the denominator of the corresponding $N$-th term in Eq. (3) reduces, producing a peak in the conductivity. The LMA oscillations are modulated by the generalized DKC effect represented by the $N$-th order Bessel function $J_N$. Thus, (3) describes a superposition of both types of 1D AMRO. In our case, data are obtained by rotating the magnetic field by angle $\theta$ from the *b*-axis (along which the heat current is applied) in the *ba*-plane (or *zy*-plane), at fixed in-plane angle $\phi = 77^0$, counted from the *c*-axis (Fig. 2c). Quantum oscillations in TEP, associated with Landau quantization are observed above *B*=8 T, the threshold field at which the $\alpha$-frequency oscillations emerge. In the present case, the energy spectrum consist of Landau levels of only the $\alpha$ orbit since the higher harmonics of the fundamental $\alpha$ orbit have not been revealed in the oscillation frequency spectrum obtained from the fast Fourier transform (FFT) analysis. The anomalously strong second harmonic, 2$\alpha$, leading to splitting of the fundamental oscillations, was found to emerge at temperatures below 2K in magnetoresistance (MR) [21-26]. The theoretical calculations were performed by using the expressions for the background TEP, Eqs.(1-5), and then summing the relation $S_{zz}^{osc} \approx \alpha_{zz}^{osc}/\sigma_{zz}$ for the oscillatory component of the TEP. The amplitude of the experimental quantum oscillations is small; the same in order as the magnitude of the non-magneto oscillatory background (see text below). This can be ascribed to the small fundamental $\alpha$ frequency of about 670 T which does not strongly enhances the effect (since the higher harmonics of the fundamental $\alpha$ orbit are absent) when performing the derivate of the DoS over $\mu$ in order to obtain the expression for the thermoelectric coefficient $\alpha_{zz}^{osc}$. A similarity between the magnetic-field dependence of the TEP and the resistance can be expected from the Boltzmann transport theory. Following the previous research on the magnetic-field dependence of the TEP for all three crystallographic directions [5], we find that in the present case $S_{zz}^b$ reflects some aspects of the in-plane components $S_{xx}^a$ and $S_{xx}^c$ suggesting that there is a mixing of these components which affects the general behavior of $S_{zz}^b$. This corroborates the fact that experimentally, there is always the possibility of random diffusion of heat along the sample due to a misalignment between the heat current and the crystal axes. Also, the total TEP may involve a mixing of contributions from several bands (in a multiband system there is a mixing between the states on the two FS sections leading to MB). For $T = 4$ K, below the transition temperature $T_p$, (Fig.1), the TEP field dependence is similar at all field directions with some features that must be addressed. TEP exhibits a minimum at $B=B_{min}$ after which it rapidly increases up to maximum at $B=B_{max}$. $B_{min}$ is angle dependent, and slightly increases with increasing angle. In contrast, $B=B_{max}$ at which the broad peak is observed in the TEP field dependence slightly decreases with increasing angle. While the minimum in the TEP becomes more broadened, the maximum is less broadened with increasing angle. This kind of behavior originates from the angular dependence of the background TEP which is very sensitive to the Fermi topology. The attenuation of the experimental observed quantum oscillations below $B_k =23$ T is due to the so-called ''kink'' field-induced phase transition. Above the ''kink field'' $B_k$, in the high-field state of CDW, where the original FS is recovered, field dependence is weak at all field directions. The suppression of the quantum oscillations at higher angles from the *b*-axis (Fig.1b and Fig. 1c) is associated with the elliptical form of the $\alpha$ orbit which becomes more elongated with tilting the field away from the *b*-axis. Due to the pronounced elongation an electron fails to perform a complete revolution during the mean free time. This is in accord with the AMRO experiments [27, 28] which have already shown that the $\alpha$ pocket is more elliptical than originally anticipated. In order to accommodate the small energy gap of ~23 meV between the q1D and q2D band, which is much smaller than band-structure predictions, the $\alpha$ pocket must become more elliptical and the q1D sheets must become more strongly warped. Although the above proposed theoretical model can account for many of the experimentally observed features in the TEP, certainly there are some differences that should be pointed out. For example, the theory yields higher amplitude of the quantum oscillation particularly above the kink field, i.e., in the high-field CDW state. A possible reason for this difference, and which has not been taken into account in the above model, could be that the CDW wave vector $\boldsymbol{Q}$ becomes field dependent in the high-field state [29].

We have also analyzed the background MTEP in order to obtain an additional insight on charge carriers involved in the transport as well as the role of the MB. The two signals were separated by removing the oscillatory MTEP by filtering out the total MTEP signal to obtain the background MTEP data. Theoretical calculations of the background MTEP were performed by using the Eqs. (1-5); the experimentally determined values valid for $\alpha$-(ET)$_2$KHg(SCN)$_4$ in its low and high field state were used. The background TEP is predominantly hole-like, i.e., positive for angles close to the $b$-axis as shown in Fig.3. As the field is rotated away from the temperature gradient direction, at angle $\theta = 45^0$, the TEP changes its sign from positive to negative at a field $B$=18 T, indicating a change in the transport from hole- to electron-like. Following the data, one can deduce that with further increasing the magnetic field tilt, the TEP will probably tend to predominantly electron-like behavior. This is, however, a reasonable conclusion since as the field approaches the plane larger number of carriers is involved in the transport. In order to simulate the background TEP at $\theta = 0^0$ and $\theta = 26^0$ we have used only the q1D expression for the conductivity while for $\theta = 45^0$ the q2D expression (Eq. 5) must be included, implying that the low-temperature TEP for angles close to the temperature gradient is determined mostly by the orbits on the new open FSs in the whole magnetic field range. Previous reports on low-temperature magnetoresistance indicate that it is determined by the open orbits only in fields below 10-15 T [30-32]. The existence of the open sheets on the FS after the reconstruction below $T_p$, is in agreement with the model previously given by Harrison *et al.* [33]. If, below $T_p$, a modified closed orbit band still remains, which can give a metallic TEP contribution, its contribution is negligible small at lower angles. However, the change in the sign, at high fields for larger field tilt seems to be connected with the formation of larger closed orbits due to the MB. This on the other hand, supports the model given by Uji *et al.* [34] according to which at increasing field, the MB through the gaps between the open and closed orbits occurs, and new, larger closed orbits with electron-like carriers can be realized. Due to the MB, larger and larger FS sections at higher fields become involved in the background and magnetic oscillations. The magnetic field applied in a general direction leads to orbits which are on a cross-section perpendicular to **B**. Bragg reflection sets the quasiparticle on a different cross section. Thus, at higher fields, the contribution from the q2D charge carriers group increases with the MB, more and more electrons are involved in the transport which explains the electron-like behavior of the TEP.

In addition to TEP we have also measured the Nernst effect, $S_{yz}^b$, for field directions $\theta = 26^0$ and $\theta = 45^0$. The total Nernst signal for $T = 4$ K is shown in Fig.4, and the corresponding background component is given in Fig.5. The Nernst effect, as TEP, also displays quantum oscillations but the Nernst response dominates by far (about 2.5 times bigger than TEP at $\theta = 26^0$ and 10 times at $\theta = 45^0$), which can be attributed to the phonon drag effect resulting from the electron-phonon coupling. Comparing the background Nernst effect with the TEP background one can see that the former has maximum approximately at the same field where the latter has minimum. The Nernst maxima are also angle dependent, increasing with increasing the angle. With increasing field, there is a decrease of the Nernst signal at all directions up to the ''kink field'' (as in the case of TEP) above which is asymptotic to non-zero angle dependent values. The background data show that the Nernst effect is not sign-change sensitive. It is possible that the MB effect is compensated with the induced transverse voltage by the longitudinal temperature gradient as the field is rotated towards the layer plane.

At temperature $T = 9$ K, just above the transition temperature $T_p$, in the metallic state of α-(ET)$_2$KHg(SCN)$_4$ there is a significant decrease of the TEP (Fig.6) and Nernst effect quantum oscillations amplitude (Fig.7). At this temperature the total TEP and Nernst effect are determined by the closed sections of the FS. While the semi-classical MR, TEP and Nernst effect, depend weakly on the temperature their oscillatory components decrease rapidly when $k_BT$ becomes of order of or greater than the distance $\hbar\omega_c$ between Landau levels, even for $k_BT<<\mu$. The factor that decreases the amplitude of the quantum oscillations is R$_T$ and for Λ>1 it falls off exponentially with temperature. For $T = 9$ K, there is a peak at $\theta \sim 30^0$ and a broad dip at $\theta \sim 49^0$, (which is close to the angles where previously the q1D AMRO dip and peak were observed for $T = 4$ K), as expected in the case of the 2D AMRO (Fig. 2b). When $\omega_c\tau\cos\theta > 1$, then the first term in Eq. (5) is dominant. However, if $\xi\tan\theta$ equals a zero of the 0th-order Bessel function, then at that angle $\sigma_{zz}^{2D}$ will be a minimum and $R(\theta)$ will be a maximum. If $\xi\tan\theta >> 1$, then the zeros occurs at angles $\theta = \theta_n^{max}$ given by $\xi\tan\theta_n^{max} = \pi(n-1/4)$, $n = 0,1,2,3...$ For $R(\theta)$ to be a minimum it should be $\theta = \theta_n^{min}$, where $\xi\tan\theta_n^{min} = \pi(n+1/4)$. According to Eqs. (1,2 and 5), the TEP is zero at angles where $R(\theta)$ is maximum or minimum as expected in case when there exist only closed FS sections. This implies that the maxima of the MR are shifted with respect to the maxima of the TEP and may give information about the transfer integral in the energy dispersion. The background TEP field dependence at these angles is shown in Fig. (8). For angles close to the *b*-axis, the TEP field dependence is sub-linear, tending to saturation at higher fields. At $\theta \sim 49^0$, as the field is oriented closely to the plane, the TEP decrease with field without saturation and is negative. In contrast, the Nernst effect, (Fig.9), is positive and increases with angle due to the induced voltage along the *y*-axis while is almost constant with increasing field.

**CONCLUSION**

In summary, we have experimentally measured the field dependence of the TEP and Nernst effect in the multiband organic conductor α-(ET)$_2$KHg(SCN)$_4$ at different field orientations and temperatures 4 K and 9 K, below and above the transition temperature $T_p$=8 K. In order to explain the results we propose a theoretical model for q1D and q2D bands which accounts for many of the features in the observed experimental results. The background TEP and Nernst effect were extracted from the total signal and its angular dependence was analyzed. The observed quantum oscillations in both thermoelectric effects originate only from the fundamental $\alpha$ frequency. The proposed theoretical model indicates that at temperatures below the transition temperature, in the CDW state, the MTEP is determined mostly by the new open Fermi sheets. Moreover, due to the small energy gap between the bands in α-(ET)$_2$KHg(SCN)$_4$, the MB plays a key role in explaining the field dependent behavior of the MTEP in the CDW state. This is in accord with the previously reported CDW scenario of the low-temperature state of α-(ET)$_2$KHg(SCN)$_4$ with imperfect nesting of the FS open sections. Nevertheless, some differences between the theory and the experiment have to be addressed. Additional research especially at lower temperatures, i.e., deep in the CDW state, is necessary to probe the above proposed theoretical model.

**ACKNOWLEDGEMENTS**

D. K. acknowledges the support from the Fulbright Program, sponsored by the U.S. Department of State and E. S. is supported in part by NSF DMR-1005293. We are very grateful to A. Kiswandhi and L. Winter for assistance with the experiments. The work was performed at the National High Magnetic Field Laboratory, supported by NSF DMR-0654118, by the State of Florida, and the DOE.

**FIGURE CAPTIONS**

**FIG. 1**. Field dependence of the total Seebeck effect signal $S_{zz}^{b}$, for $T = 4$ K in the CDW state at field tilted for a) $\theta = 0^{0}$, b) $26^{0}$ and c) $45^{0}$ from the *b*-axis.

**FIG. 2.** AMRO for a) $T = 4$ K, $B = 15$ T and b) $T = 9$ K, $B = 25$ T. (c) Conventional definition of the magnetic field orientation by the tilt angle $\theta$ and the azimuthal angle $\phi$. The temperature gradient $\nabla T$ is applied along the *b*-axis.

**FIG. 3**. Background thermopower of $S_{zz}^{b}$, for $T = 4$ K in the CDW state at the same field orientations obtained from Fig. 1 by filtering out the quantum oscillation component.

**FIG. 4**. Field dependence of the total Nernst effect signal $S_{yz}^{b}$, for $T = 4$ K in the CDW state at a) $\theta = 26^{0}$ and b) $45^{0}$ from the *b*-axis.

**FIG. 5**. Background Nernst effect $S_{yz}^{b}$, for $T = 4$ K in the CDW state at the same field orientations, obtained from Fig. 4 by filtering out the quantum oscillation component.

**FIG. 6**. Total Seebeck effect $S_{zz}^{b}$, for $T = 9$ K in the metallic state at field tilted for a) $\theta = 0^{0}$, b) $30^{0}$ and c) $49^{0}$ from the *b*-axis.

**FIG. 7**. Total Nernst effect $S_{yz}^{b}$, for $T = 9$ K in the metallic state at field tilted for a) $\theta = 30^{0}$ and b) $49^{0}$ from the *b*-axis.

**FIG. 8**. Background thermopower of $S_{zz}^{b}$, for $T = 9$ K in the metallic state at the same field orientations as in Fig. 6.

**FIG. 9**. Background Nernst effect of $S_{yz}^{b}$, for $T = 9$ K in the metallic state at the same field orientations as in Fig. 7.

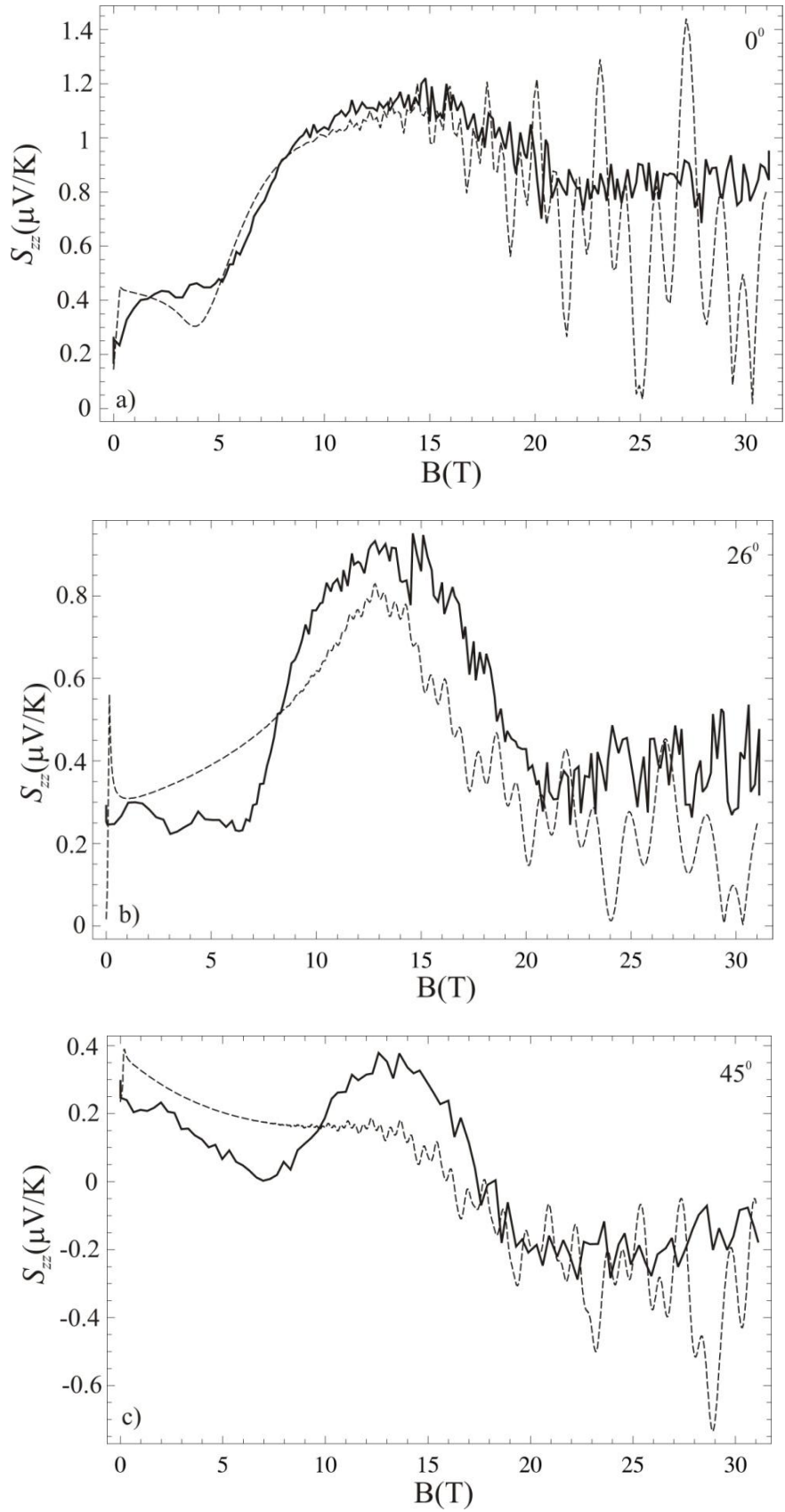

**FIG. 1**

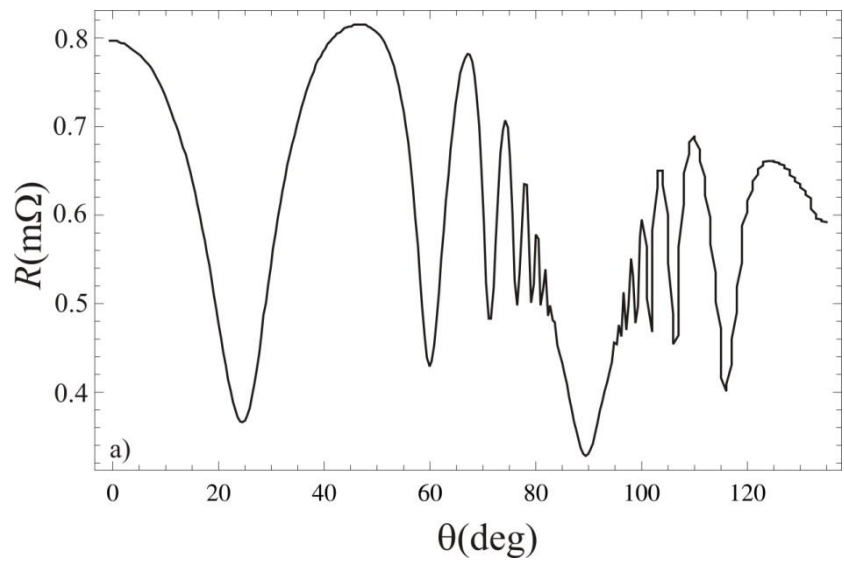

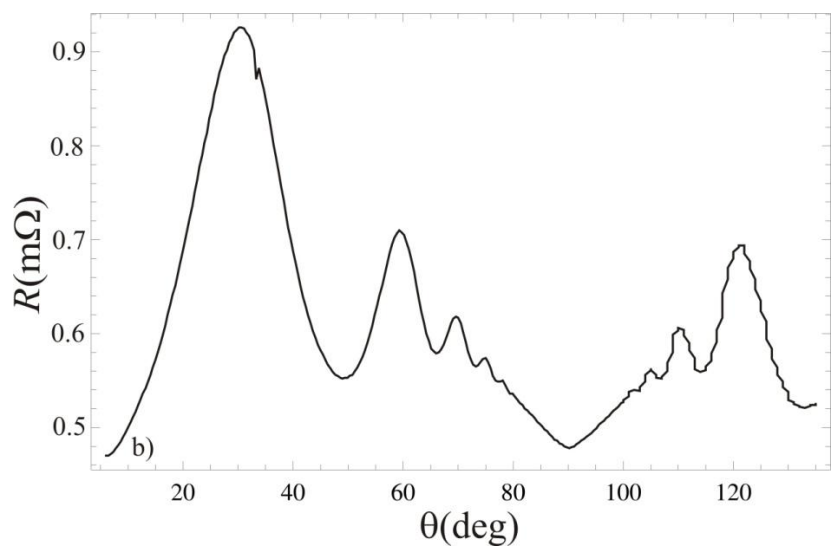

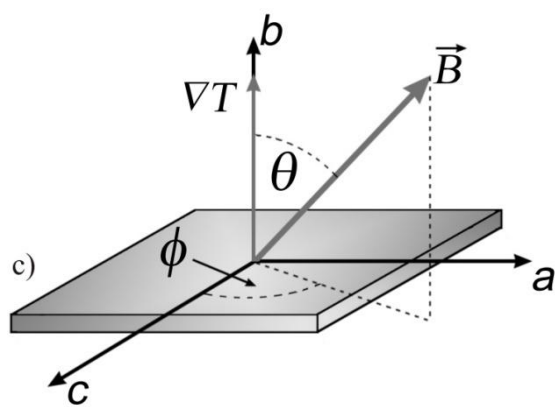

**FIG. 2**

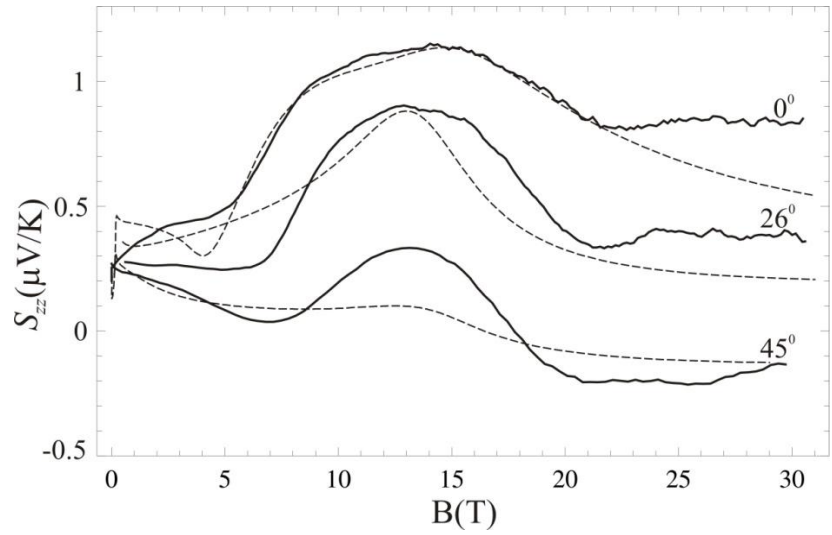

**FIG. 3**

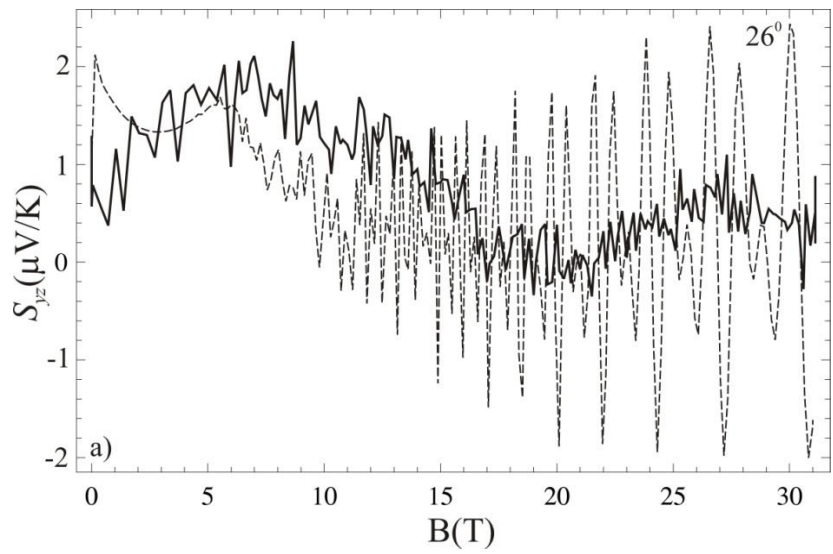

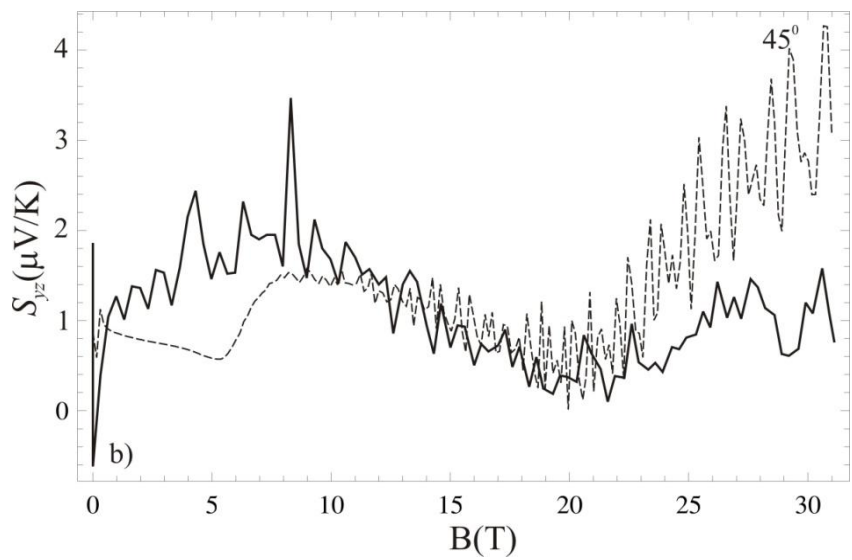

**FIG. 4**

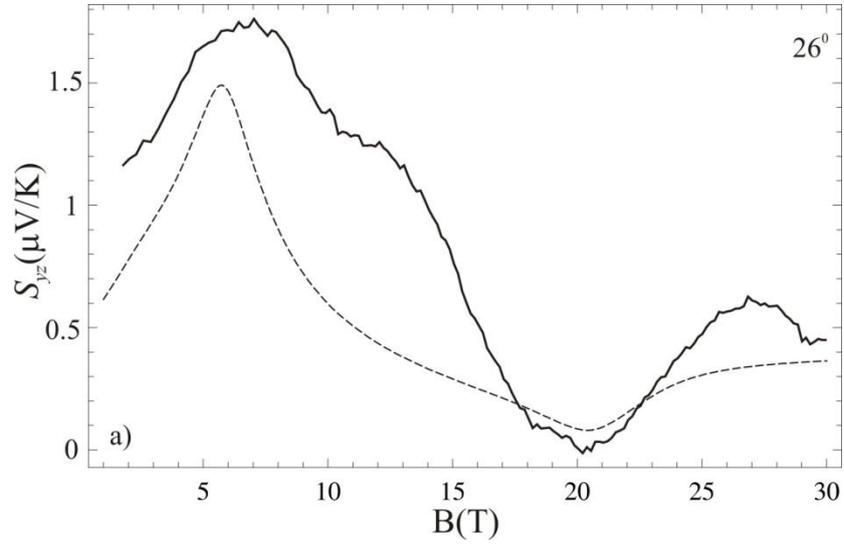

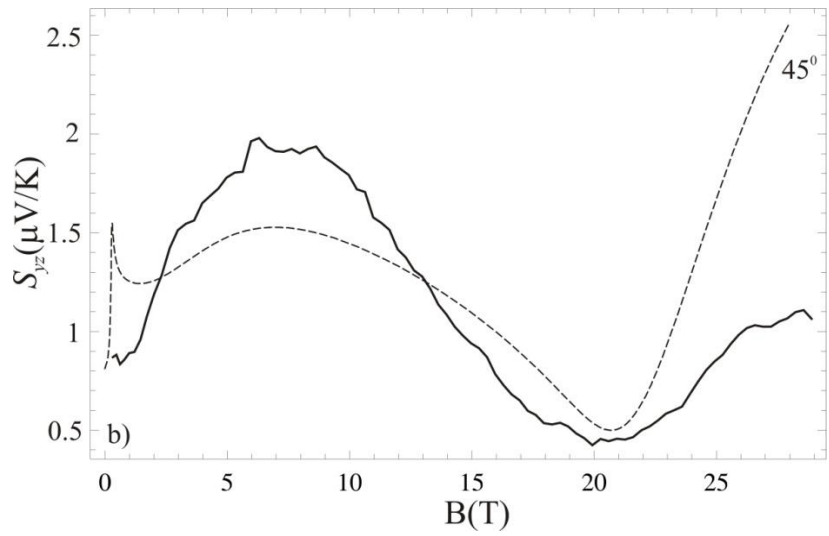

**FIG. 5**

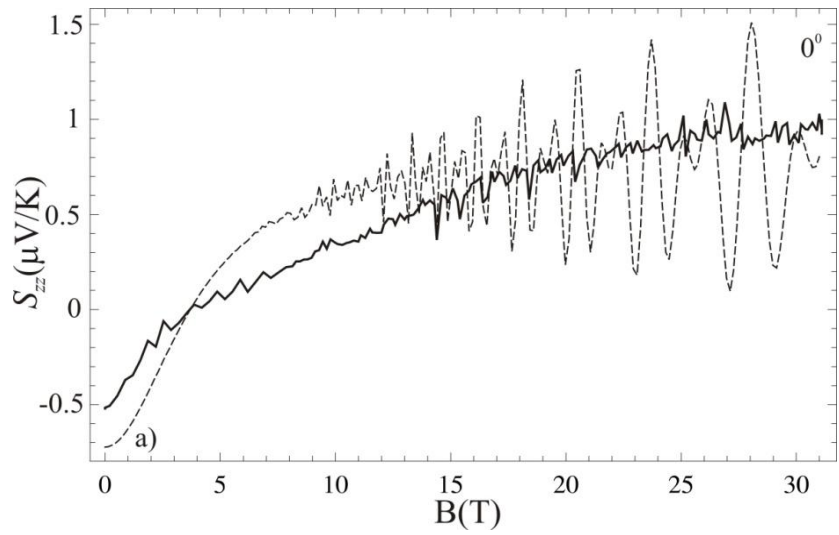

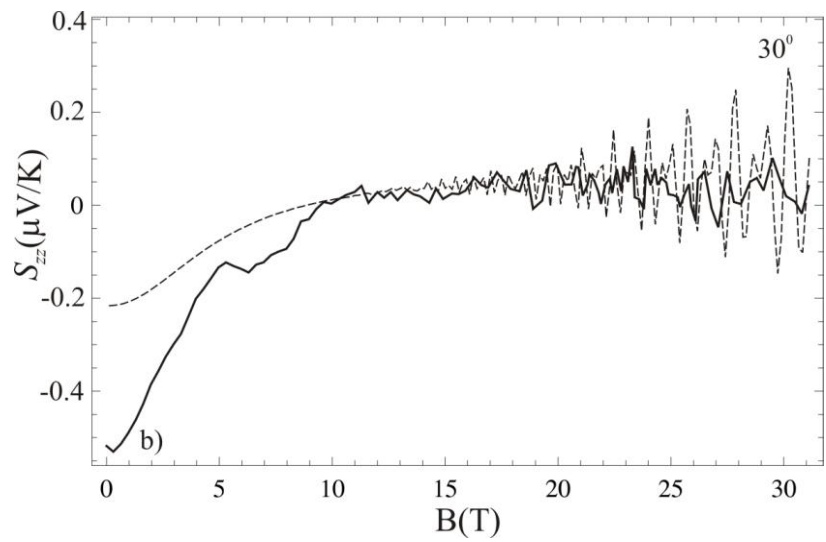

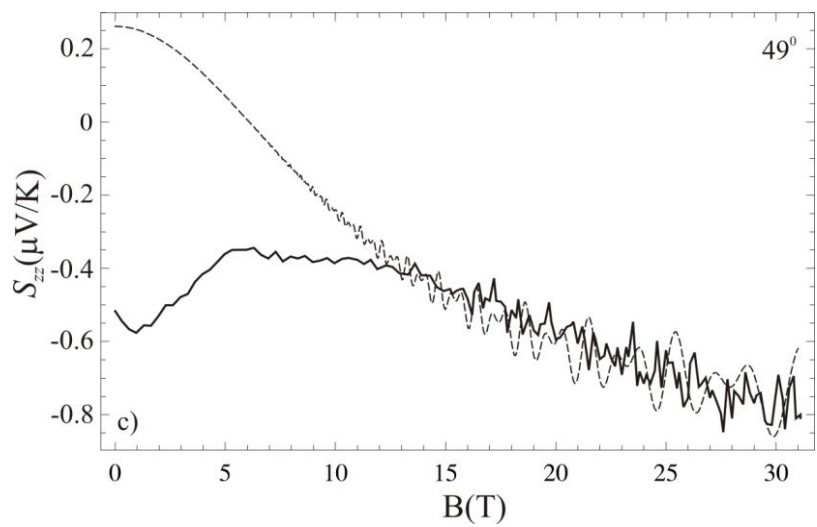

**FIG. 6**

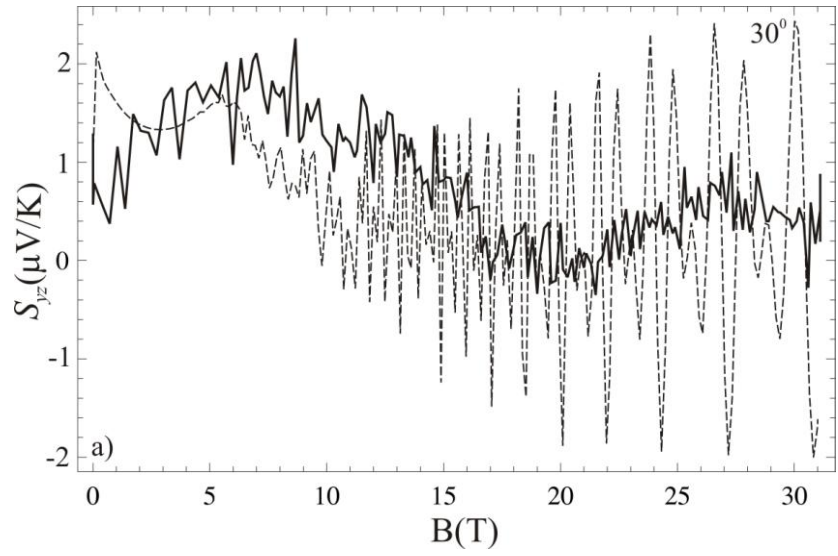

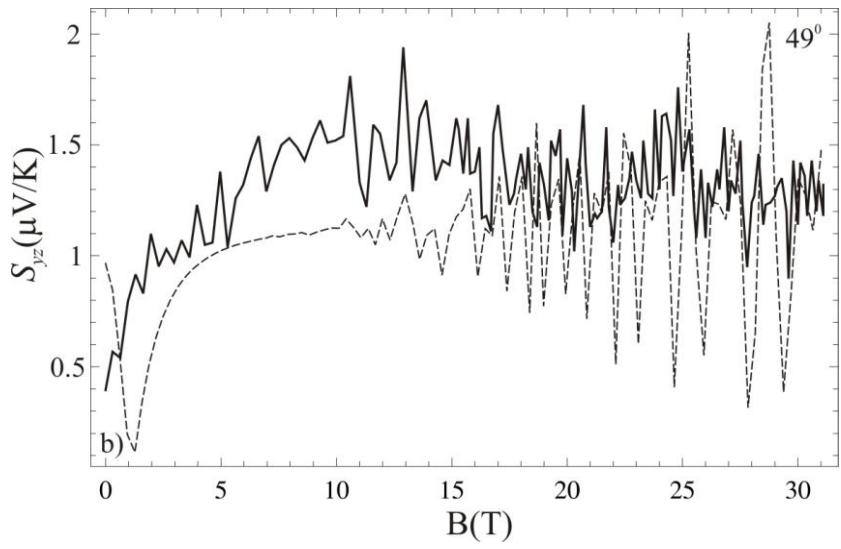

**FIG. 7**

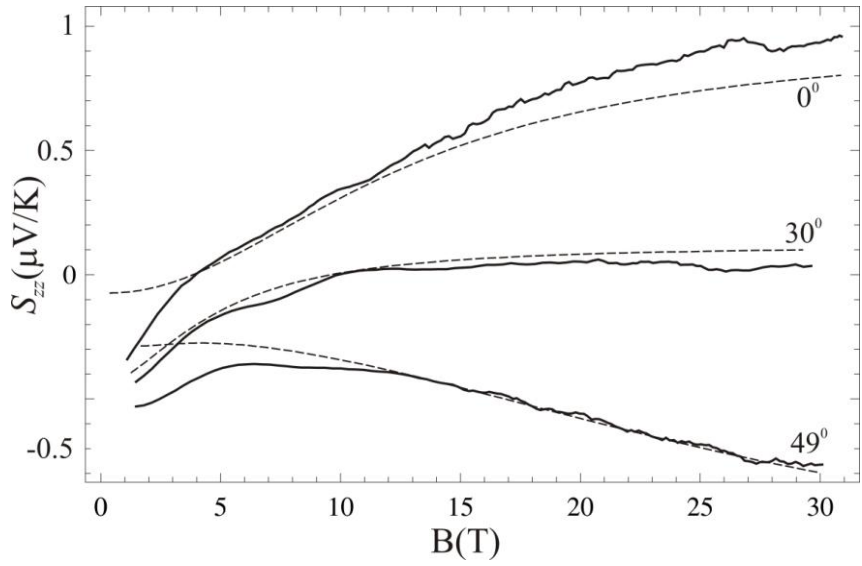

**FIG. 8**

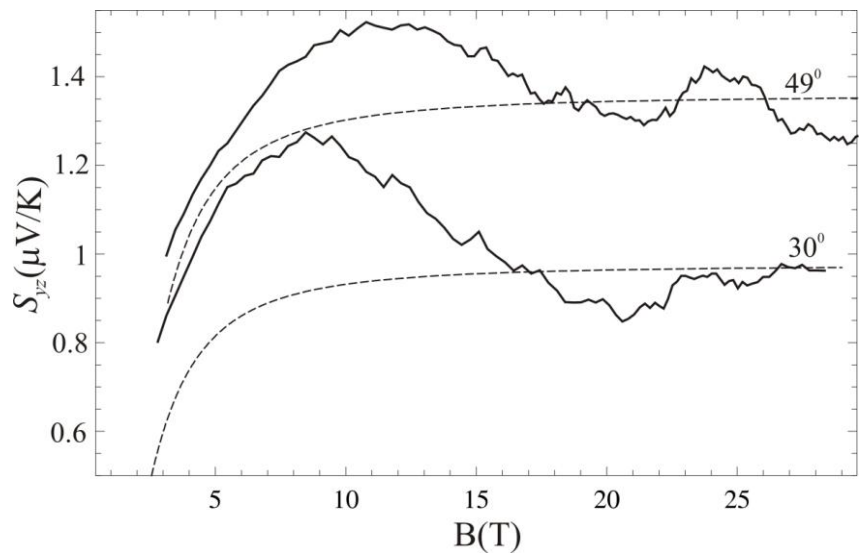

**FIG. 9**